\documentclass[aps,pra,twocolumn,groupedaddress,draft,showpacs]{revtex4}
\begin{document}

\title{Density-functional calculation of ionization energies of 
current-carrying atomic states}
\author{E. Orestes}
\author{T. Marcasso}
\author{K. Capelle}
\email{capelle@if.sc.usp.br}
\affiliation{Departamento de Qu\'{\i}mica e F\'{\i}sica Molecular\\
Instituto de Qu\'{\i}mica de S\~ao Carlos\\
Universidade de S\~ao Paulo\\
Caixa Postal 780, S\~ao Carlos, 13560-970 SP, Brazil}
\date{\today}

\begin{abstract}
Current-density-functional theory is used to calculate ionization energies 
of current-carrying atomic states. A perturbative approximation to full 
current-density-functional theory is implemented for the first time, and 
found to be numerically feasible. Different parametrizations for the
current-dependence of the density functional are critically compared.
Orbital currents in open-shell atoms turn out to produce a small shift in 
the ionization energies. We find that modern density functionals
have reached an accuracy at which small current-related terms appearing
in open-shell configurations are not negligible anymore compared to
the remaining difference to experiment.
\end{abstract}

% 31.15.Ew Density-functional theory
% 32.30.-r Atomic spectra
% 31.30.-i Corrections to electronic structure
% 32.60.+i Zeeman and Stark effects

\maketitle

\newcommand{\be}{\begin{equation}}
\newcommand{\ee}{\end{equation}}
\newcommand{\bea}{\begin{eqnarray}}
\newcommand{\eea}{\end{eqnarray}}
\newcommand{\bi}{\bibitem}

\renewcommand{\r}{({\bf r})}
\newcommand{\rp}{({\bf r'})}

\newcommand{\ua}{\uparrow}
\newcommand{\da}{\downarrow}
\newcommand{\la}{\langle}
\newcommand{\ra}{\rangle}
\newcommand{\dg}{\dagger}

\section{\label{intro}Introduction}

Density-functional theory (DFT) \cite{hk} is based on identification of the 
single-body charge density $n\r$ as key variable in terms of which all 
ground-state 
observables of an interacting many-electron system can be described
\cite{dftbook,parryang}. Although $n\r$ is in principle sufficient for this 
purpose, in practice it often turns out to be advantageous to employ
additional variables. The most commonly employed such additional variable is
the spin magnetization ${\bf m}\r$ (or the spin-resolved charge densities
$n_\ua\r$ and $n_\da\r$), leading to spin-density functional theory (SDFT)
\cite{dftbook,parryang,gunlund}. Other choices have occasionally been found 
useful, e.g. in solid-state physics \cite{scdft,sdwdft}.  

In the present paper we are interested in orbital magnetism produced by
currents forming in open-shell atoms. The current density seems a useful 
extra variable in this situation. In the absence of external magnetic fields
ground-state currents can in principle be calculated also by means of (S)DFT. 
However, in practice the calculation of orbital currents and their effects 
on observables is impossible in standard formulations of DFT and SDFT, because 
no explicit prescription for calculating the true (many-body) current density 
is known in these theories. Of course, one can always calculate the current 
arising from the single-particle Kohn-Sham (KS) orbitals of SDFT or DFT, but 
these orbitals are constrained only to reproduce the correct charge (and spin)
densities of the interacting many-body system, and there is no guarantee
that the current calculated from them bears any relation to the true current.

Orbital magnetism is thus basically out of reach of conventional DFT.
In view of the importance of currents, appearing either spontaneously or 
induced by external magnetic fields in a wide variety of many-body systems,
such as atoms and molecules with open shells, nuclear-magnetic resonance, 
cyclotron resonance, superconductivity, and magnetism of transition metal 
and rare-earth compounds, it is thus desirable to develop a DFT-based
approach that allows to directly address the effects of orbital currents.
In the present paper we explore one such formulation of DFT, namely
nonrelativistic current-density functional theory (CDFT) \cite{vr1,vr2,vr3}.
Relativistic DFT \cite{strange,rajagopal,macdonald} in principle also provides
explicit information on the current, but standard implementations of it are
formulated in a spin-only version, which prohibits extraction of information
on the currents. Furthermore, the formalism of relativistic DFT is considerably
more complicated than that of SDFT.

CDFT is formulated in terms of the charge density $n\r$ and the
nonrelativistic paramagnetic current density vector
\bea
{\bf j}_p\r= \sum_k {\bf j}_{p,k}\r 
\nonumber \\
=\frac{\hbar}{2mi}\sum_k
\left[\psi_k^*\r\nabla\psi_k\r-(\nabla\psi_k^*\r)\psi_k\r\right].
\label{jcdft}
\eea
This current is to be calculated from the CDFT Kohn-Sham equations 
\cite{vr1,vr2,vr3}
\be
\left[\frac{1}{2m}\left(\frac{\hbar}{i}\nabla-\frac{q}{c}{\bf A}_s\r \right)^2
+V_s^c\r \right]\psi_k\r
=\epsilon_k^c \psi_k\r,
\label{cdftks}
\ee
where an upper index `$c$' denotes CDFT,
\be
V_s^c\r= v_s^c\r + \frac{q^2}{2mc^2}
\left({\bf A}_{ext}\r^2 - {\bf A}_s\r^2 \right),
\label{vcdft}
\ee
\be
v_s^c\r = v_{ext}\r+v_H\r+v^c_{xc}\r
\ee
and
\be
{\bf A}_s\r={\bf A}_{ext}\r + {\bf A}_{xc}\r.
\label{acdft}
\ee
Here $v_{ext}$ and ${\bf A}_{ext}$ are external static electric and magnetic
potentials, $v_H$ is, as usual, the Hartree potential, and $v^c_{xc}$ and
${\bf A}_{xc}$ are the exchange-correlation ($xc$)
scalar and vector potentals of CDFT, respectively \cite{vr1,vr2,vr3}.
Gauge invariance of CDFT requires that the $xc$ energy $E_{xc}[n,{\bf j}_p]$
depends on the current only through the so called vorticity
\cite{vr1,vr2,vr3}
\be
\nu[n,{\bf j}_p]\r=\nabla\times\frac{{\bf j}_p\r}{n\r},
\label{pvort}
\ee
i.e., is of the form $E_{xc}[n,{\bf j}_p]=\bar{E}_{xc}[n,\nu]$.
This dependence provides a useful constraint on approximate CDFT functionals.  

Although CDFT formally solves the problem how to obtain current-related
information from DFT, many practical questions remain. One is, clearly, the
construction of approximate current-density functionals. A brief summary of 
progress in this area is given in Sec.~\ref{functional}. Another is the
actual implementation of CDFT. In practice, a fully self-consistent solution
of the CDFT equations is still quite demanding.
Furthermore, in many interesting situations the effect of orbital
magnetism, while important for a qualitative and quantitative understanding,
is relatively small, so that a fully self-consistent treatment of the
orbital degrees of freedom is not always required.
The question thus arises if one can put the insights, gained from the formalism
of full-fledged CDFT, at work within {\em conventional} DFT, in order to
achieve an improved description of orbital magnetism via a self-consistent
solution of the widely implemented traditional Kohn-Sham equation.

A simple answer to this question was given by one of us in Ref.~\cite{jpert},
by pointing out that the CDFT Kohn-Sham equations can be written in the
form of the DFT ones plus a remainder that depends explicitly on the
$xc$ and external vector potentials. In the absence of external magnetic fields
and for not too large $xc$ vector potentials it then suggests itself to use
low-order perturbation theory in order to describe the CDFT modifications
to the DFT equations. This idea has been worked out in Ref.~\cite{jpert},
where explicit expressions for the CDFT corrections to a number of important
DFT quantities were given. The resulting approach is labeled `perturbative
CDFT', or simply pCDFT.
A simple example of pCDFT expressions is the correction to the DFT eigenvalue
spectrum $\epsilon_k^d$, which can be cast in the form \cite{jpert}
\be
\epsilon_k^c =
\epsilon_k^d - \frac{q}{c}  \int d^3r \, {\bf j}^{KS}_{p,k}\r \cdot
{\bf A}_{xc}[n,{\bf j}^{KS}_{p}]\r.
\label{epspert}
\ee

\section{\label{functional}Current-dependent local-density approximation}

In the present paper we apply pCDFT to a study of the effect of currents in 
open shells on atomic ionization energies. To this end, we require an explicit 
expression for the $xc$ energy in the presence of orbital currents.
In Refs.~\cite{vr1,vr2} Vignale and Rasolt proposed an extension of the
local-density approximation (LDA) of ordinary, charge-only, DFT to the 
case of CDFT. Their functional takes the form
\be
E_{xc}[n,{\bf j}_p] = 
\int d^3r \, n\r \bar{e}_{xc}\left(n,\nu\right),
\label{linres}
\ee
where
\be
\bar{e}_{xc}\left(n,\nu\right) = e_{xc}(n,0)+\frac{m k_F^0}{24
\pi^2}\left(\frac{\chi_L}{ \chi_L^0}-1\right)
\frac{|\nu\r|^2}{n\r}.
\label{ldakernel}
\ee
Here $k_F^0$ is the Fermi wave vector of a non-interacting electron gas,
related to the density via $k_F^0=(3\pi^2 n)^{1/3}$. The function
$e_{xc}(n,0)$ is the exchange-correlation energy per particle
in the absence of external magnetic fields, and can be approximated, e.g., 
by the usual LDA or any of the available generalized-gradient approximations 
(GGA's).  The second term on the right-hand side of
Eq.~(\ref{ldakernel}) is a weak-field (linear response) expression for the 
current-dependent part of the functional.
The functional defined by the preceeding equations has been used in 
several CDFT calculations \cite{ebert,handy1,handy2,handy3} and should be 
good enough for a first orientation about the size and nature of 
current-related phenomena.

Many-body effects enter this functional via 
the ratio of the orbital susceptibilities of the interacting and the
noninteracting electron gas, $s:=\chi_L/\chi_L^0$. This ratio has been 
calculated numerically by Vignale, Rasolt and Geldart \cite{vrg} for 10 
values of the dimensionless density parameter $r_s$, which is related to 
the density by $n =3/4 \pi (r_s a_0)^3$, where $a_0$ is the Bohr radius. 
To utilize these results in CDFT, Lee, Colwell and Handy (LCH) proposed 
the expression \cite{handy2,handy3}
\be
s_{LCH}(r_s)=(1.0 + 0.028 r_s) \exp{(-0.042 r_s)}
\label{lhcfit}
\ee
as a convenient and accurate analytical interpolation through the numerical
data of Ref.~\cite{vrg}.
In the original reference \cite{handy2} this fit is claimed to have an rms
error of $\sim 1.5\times 10^{-3}$. This appears to be a misprint in 
\cite{handy2}, because on redoing the calculation we find that over 
the 10 data points provided by Vignale, Rasolt and Geldart \cite{vrg} the
rms error of expression (\ref{lhcfit}) is a little bigger, namely
$\sim 1.9\times 10^{-3}$. We have also constructed two alternative 
interpolations through the data of Ref.~\cite{vrg}. With the same number of 
fitting parameters (three) as in the LCH fit we find that the expression
\be
s_{3}(r_s)= 0.9956-0.01254 r_s-0.0002955 r_s^2
\label{3termfit}
\ee
has an rms error of $\sim 1.2\times 10^{-3}$, while the 5-term expression 
\bea
s_{5}(r_s)=1.1038-0.4990 r_s^{1/3}+0.4423 \sqrt{r_s}
\nonumber \\ -0.06696 r_s  +0.0008432 r_s^2
\label{5termfit}
\eea
has an rms error of only $\sim 2.1\times 10^{-4}$ over the same 10 data points. 
Over the range of values of $r_s$ spanned by these data ($1\ldots 10$) the 
latter expression should thus be prefered, compared to (\ref{lhcfit}) or 
(\ref{3termfit}).
In the calculations presented below the current flows mainly in a region in
which $0.1 < r_s < 5$. 
We have therefore performed all calculations once with the LCH fit 
(\ref{lhcfit}) and once with the above 5-term fit (\ref{5termfit}).
As will be seen below, the differences between both sets of results
can be considerable. Since our calculations are self-consistent only 
with respect to the charge density, but perturbative with respect to 
the current density, we expect, however, that it is only the order of 
magnitude of the results that is quantitatively reliable, and on this 
order of magnitude both parametrizations employed agree consistently.

The LCH expression and both alternative fits differ
markedly for $r_s >12$, i.e., in the extreme low-density limit. For the
present calculation this range is less important, but in view of potential
future applications of CDFT in the low-density regime it should be noted
that the numerical data of Ref.~\cite{vrg} do not constrain the various
fits in that regime.
In the opposite limit, $r_s\to 0$, the five-term fit (\ref{5termfit}),
although on average significantly more accurate than the LCH expression, 
does not correctly recover the value at $r_s=0$, whereas the LCH fit correctly 
yields $s_{LCH}(0)=1$.
However, the limit $r_s\to 0$ corresponds to infinite density, and is thus
rather unimportant for typical atomic physics applications. Moreover, in this 
limit the asymptotically exact expression \cite{vrg}
\be
s_{r_s\to 0}(r_s)
= 1+\frac{\alpha}{6\pi}r_s\ln r_s + 0.08483 \frac{\alpha r_s}{\pi}
+ O(r_s^2),
\label{rsto0}
\ee
where $\alpha=(4/9 \pi)^{1/3}$, is available, so that there is no need for data
fitting at all in this limit.

In addition to Eq.~(\ref{linres}) with (\ref{ldakernel}) several other
CDFT functionals have been proposed, but none seems suitable for our purposes.
The approach of Ref.~\cite{handyfctnl} has not yet led to an explicit
expression for the current dependence of ${\bf A}_{xc}[n,{\bf j}_p]$.
The functional of Ref.~\cite{skudlarski} displays quantum oscillations
arising from Landau-level filling in the electron gas, and is thus suitable
only for extended systems. The functionals of Refs.~\cite{2dim,pade},
on the other hand, were designed specifically for two-dimensional systems
in the quantum Hall regime. The expression of Ref.~\cite{becke}, finally,
is not a vorticity functional. We will return to this last functional
in our discussion of the results in Table~\ref{table2}, below.
Given the scarcity of suitable functionals, the proposal of Ref.~\cite{cdftlett}
to generate a CDFT functional by means of a set of integral transformations
from an input SDFT functional may prove useful in the future. In the present 
work, however, we restrict attention to the simple linear-response LDA defined 
by Eqs.~(\ref{linres}) and (\ref{ldakernel}).

\section{\label{results}Numerical results for ionization energies of
current-carrying atomic states}

After this preparatory discussion of the CDFT functional, we now return to
the question of CDFT shifts with respect to the DFT eigenvalues. In general
the eigenvalues obtained from solution of the KS equations have no rigorous
physical meaning, although they can bear a semiquantitative relationship
to the true energy spectrum \cite{kohnbeckeparr}. One exception to this rule
is the highest occupied eigenvalue of the KS spectrum, which is known to be
the negative of the system's ionization energy, and as such can be compared
directly to experiment. Although once the system is prepared in a
current-carrying state there are, in general, CDFT shifts to all KS
eigenvalues, we therefore focus in the present paper on the highest occupied
one, since it is most tightly connected with experiment.

\begin{table*}
\caption{\label{table1} Current-induced changes in the ionization energies
of atoms with open $p$ ($B$ to $Cl$) and $d$ ($Sc,Y$) shells.
First column: atom.
Second column: zero-current ionization energy calculated within LDA-SIC-KLI,
from \protect\cite{krieger}.
Third column: experimental ionization energies, from \protect\cite{nist}.
Fourth column: current-carrying single-particle state considered here.
Fifth column: pCDFT correction in LDA, using the LCH expression (\ref{lhcfit})
for the susceptibilities.
Sixth column: pCDFT correction in LDA, using the present expression
(\ref{5termfit}) for the susceptibilities. All values in $eV$.
}

\begin{ruledtabular}
\begin{tabular}{ccclll}
& I & I & cc sp state(s) & -$\Delta \epsilon^{pCDFT}$ & -$\Delta \epsilon^{pCDFT}$\\
&LDA-SIC-KLI & exptl. & & with (\ref{lhcfit})& with (\ref{5termfit})\\
\hline
B  &8.316&8.2980 & $m=1$ & 0.072&0.056   \\
C  &11.60&11.2603 & $m_1=1,m_2=0$&0.045 &0.051   \\
   &&& $m_1=1,m_2=1$&0.18 &0.20   \\
N  &14.95&14.5341 & $m_1=1,m_2=0,m_3=0$&0.034 &0.047   \\
   &&& $m_1=1,m_2=1,m_3=0$& 0.14 &0.19   \\
O  &14.33&13.6181 & $m_1=1,m_2=1,m_3=0,m_4=0$&0.11 &0.11   \\
   &&& $m_1=1,m_2=1,m_3=0,m_4=-1$&0.027 &0.027   \\
F  &18.61&17.4228 & $m_1=1,m_2=1,m_3=0,m_4=0,m_4=-1$&0.023 &0.040  \\
Al &5.570&5.9858 & $m=1$ &0.049 &0.030  \\
Si &7.804&8.1517 & $m_1=1,m_2=0$&0.022 &0.018  \\
   &&& $m_1=1,m_2=1$&0.089 &0.073 \\
P  &10.07&10.4867 & $m_1=1,m_2=0,m_3=0$&0.015 &0.014  \\
   &&& $m_1=1,m_2=1,m_3=0$& 0.059& 0.057\\
S  &10.41&10.3600 & $m_1=1,m_2=1,m_3=0,m_4=0$&0.044 &0.044 \\
   &&& $m_1=1,m_2=1,m_3=0,m_4=-1$&0.011 &0.011  \\
Cl &13.08&12.9676 & $m_1=1,m_2=1,m_3=0,m_4=0,m_4=-1$&0.0091 &0.011  \\
Sc & - &6.5615 & $m=1$& 0.036 &0.043  \\
   &&  & $m=2$& 0.037 &0.044  \\
Y  & - &6.2171 & $m=1$& 0.035 &0.035  \\
   &&  & $m=2$& 0.036 &0.037
\end{tabular}
\end{ruledtabular}
\end{table*}

In comparisons of experimental ionization energies with KS eigenvalues
obtained within the local-density approximation (LDA) it typically
turns out that the LDA eigenvalues are significantly off.
The origin of this problem is known to be the asymptotic behaviour of the
LDA $xc$ potential, which decays too rapidly, and leads to too weak
binding of the outermost electron. Self-interaction corrections (SIC)
\cite{pz}, which correct the wrong LDA asymptotics,
significantly improve on the LDA values for ionization energies, and
typically are quite close to experimental data \cite{krieger}.
Once a self-interaction correction has been applied, and the main error
of the LDA ionization energies removed, one can consider the
effect of additional small corrections, such as the effects of orbital
currents. Such current-dependent corrections to the ionization energy can
become important, e.g., when ionization (or transfer of electrons during
formation of chemical bonds) takes place in the presence of external magnetic
fields, since such fields polarize the atom and can give rise to orbital 
currents. 

It is in this situation, ionization or electron transfer in the presence of 
static external magnetic fields, where pCDFT calculations of current-induced
shifts of the ionization energies are directly applicable.
Within the context of density-functional theory, the possibility of small
current-induced shifts in the ionization energies is also relevant for the 
calculation of excited states from time-dependent DFT (TDDFT). A recent 
systematic investigation \cite{tddft} of sources of error in excitation
energies calculated from TDDFT concludes that the most important ingredient
in such calculations is the ground-state $xc$ potential used to generate the 
Kohn-Sham response function, and that `the most important requirement for 
such a potential would be that its highest occupied eigenvalue reproduces 
the experimental ionization potential as closely as possible' 
(cf. Sec.~6 of Ref.~\cite{tddft}).
TDDFT calculations of excitations from open shells are thus expected to
sensitively depend on current-induced shifts of the ionization energies,
if the excitation takes place in the presence of external magnetic fields.

  A separate issue is whether a CDFT calculation of the type presented here 
can also be useful in the absence of any external magnetic field. An
argument for such utility could run along the following lines: The negative of
the highest occupied KS eigenvalue of DFT gives the true ionization energy.
The negative of the highest occupied eigenvalue of CDFT also gives this energy.
In the absence of external magnetic fields both values must thus be identical, 
if one works with the exact functionals. In practise, of course, we do not have
the exact functionals available, and it becomes a meaningful question to ask 
which of the two approximate eigenvalues is closer to the experimental energy. 
Our original expectation (not consistently confirmed by the numerical data
shown below) was that for atoms whose many-body ground state has a nonzero
value of the total angular momentum quantum number $L$ the CDFT value would
be better than the DFT one, if the CDFT-KS calculation is performed for a
system prepared in a state with an orbital current.

Motivated by these considerations we have numerically calculated the pCDFT 
correction to the highest occupied KS eigenvalue for a series of atoms with 
unfilled $p$ and $d$ shells, prepared in a current-carrying state.
Energies and radial wave functions of the unperturbed system were obtained 
numerically from a standard spherically averaged DFT calculation, using the 
basis-set-free program 
{\it opmks} \cite{opmks}, both within the LDA and the BLYP GGA \cite{blyp}.
The full single-particle orbital for a current-carrying state was then
obtained by multiplying the radial wave function with a spherical harmonic
corresponding to a definite value of the magnetic quantum number $m$.
Such states carry a current proportional to $m$. This procedure defines
how we prepare a current-carrying state in the KS system. Experimentally,
the preparation of an open-shell atom in a current-carrying state is achieved
by applying suitable external magnetic or electric fields, as in the Zeeman 
and Stark effects \cite{footnote1}.

In Table \ref{table1} the results of LDA-SIC and pCDFT calculations
of atomic ionization energies are listed for a number of current-carrying
states. It is interesting to see that the differences between values
for $\Delta \epsilon^{pCDFT}$  obtained with 
both parametrizations of the susceptibility data used in this work, 
Eq.~(\ref{lhcfit}) and Eq.~(\ref{5termfit}), can be substantial, although the 
order of magnitude predicted by both is the same. On the other hand, it makes 
little difference for the size of $\Delta\epsilon^{pCDFT}$ whether the 
unperturbed orbitals used for calculating ${\bf j}_{p,k}\r$ and 
${\bf A}_{xc}\r$ are obtained from LDA or GGA. The values listed in 
Table~\ref{table1} were obtained with LDA. For comparison we also performed 
calculations using the BLYP GGA \cite{blyp}, but the resulting changes 
in the pCDFT corrections are consistently smaller than the ones arising from 
changing the parametrization of the current-dependent part of the functional 
from (\ref{lhcfit}) to (\ref{5termfit}).
On the other hand, depending on the particular occupation of $m$-substates 
in the noninteracting system
the size of both the current and the resulting energy shift can vary 
considerably. To explore this variation Table~\ref{table1} lists values for 
more than one occupation for most atoms.

Comparison of the experimental ionization energies listed in Table \ref{table1}
with the sum of LDA-SIC-KLI values and the pCDFT correction shows that 
pCDFT slightly improves agreement with experiment for several 
second-row elements, but worsens it for the first-row ones.
The data are thus inconclusive as to whether pCDFT represents a true
improvement on calculations of ionization energies that neglect currents.
This is similar to the situation encountered by LCH in the CDFT calculation 
of nuclear shielding tensors, in Ref.~\cite{handy3}, where the same functional 
(\ref{ldakernel}) as here was employed and CDFT was found to yield rather small
shifts that did not systematically improve agreement with experiment.

In Ref.~\cite{handy3} this was attributed to deficiencies in the 
approximation for the density functional. Although the same could be true 
here, too, it is in our opinion more likely that the inconclusive
comparison between LDA-SIC-KLI+pCDFT ionization energies with experiment is
due to the fact that while the pCDFT values were obtained by explicitly 
assuming a current-carrying configuration in the open shell, the experimental 
data refer to configurations with on average equal population of all 
$(2L+1)(2S+1)$ degenerate substates belonging to the $^{2S+1}L$
ground-state term of the atom. Such an average eliminates the
orbital currents. A direct comparison of the sum of the LDA-SIC-KLI values 
and the pCDFT ones with available experimental data is thus not necessarily 
meaningful. Rather, experiments in the presence of suitable external fields,
selectively populating states with a nonzero orbital current, are called for. 

\begin{table}
\caption{\label{table2}                                                         Comparison of order of magnitude of current-related effects in open
shells with accuracy of LDA-SIC calculations.
First column: atom considered.
Second column: absolute difference between LDA-SIC-KLI data
\protect\cite{krieger} and experimental result \protect\cite{nist}.
Third column: current-induced shift of ground-state energies, as
calculated from fourth and fifth columns of Table~2 of
Ref.~\protect\cite{becke}.
Fourth column: current-related contribution to ionization energy of
current-carrying states, taken from the last column of Table~\ref{table1}.
(For atoms for which more than one configuration is listed in Table
\ref{table1} we have taken the one with the more negative
$\Delta \epsilon^{pCDFT}$ for the present comparison.)
Note that the numbers in columns two, three and four measure different things,
and should not expected to be identical. As discussed in the main text, the
fact to note is that they are of similar magnitude, showing that
current-related phenomena in open-shell atoms (columns three and four) lead
to energy shifts that are comparable to the precision of modern density
functionals (column two).
All values in $eV$.}
\begin{ruledtabular}
\begin{tabular}{rlll}
& deviation of & $-\Delta E({\bf j})$ & $-\Delta \epsilon^{pCDFT}$ \\
& LDA-SIC-KLI \protect\cite{krieger} & from Ref.~\protect\cite{becke} & with (\ref{5termfit})\\
& from experiment \protect\cite{nist} & & \\
\hline
B&0.018&0.14&0.056\\
C&0.34&0.15&0.20\\
N&0.42&-&0.19\\
O&0.71&0.23&0.11\\
F&1.2&0.24&0.040\\
Al&0.42&0.075&0.030\\
Si&0.35&0.069&0.073\\
P&0.42&-&0.057\\
S&0.050&0.092&0.044\\
Cl&0.11&0.085&0.011
\end{tabular}
\end{ruledtabular}
\end{table}

Even in the absence of such experiments, it is, however, still possible 
to compare the size of the pCDFT shifts in Table \ref{table1} with
other calculations of the effects of orbital currents in open-shell atoms
and with the accuracy of present-day DFT calculations. Such a comparison is 
presented in Table \ref{table2}, in which we compare the order of magnitude 
of the pCDFT shifts in Table \ref{table1} with the deviation 
of LDA-SIC calculations from experiment, and with results from another recent 
calculation of effects of orbital currents in open-shell atoms.
From the numbers in Table \ref{table2} it is obvious that once the effect of
the self-interaction corrections has been taken into account, the order of
magnitude of the effect of currents in the open shell is comparable
to the remaining difference between the LDA-SIC data and experiment.
As explained above, this does not mean that agreement with experiment is 
necessarily improved by simply adding the two contributions, but it implies 
that further refinement of density functionals for the calculation of the
electronic structure of atoms must take the possibility of curent-dependent 
energy shifts into account \cite{tddft,jcpbaerends,becke}.

In recent work of Becke \cite{becke} the effect of orbital currents in
open shells is studied from a different point of view. Instead of
explicitly preparing the atom in a symmetry-broken current-carrying state 
and studying the resulting shifts of single-body energies, as we do here, 
he considers the total ground-state energies calculated from different 
but symmetry-equivalent configurations, which should be degenerate.
Nevertheless, approximate density functionals can give rise to an artificial 
breaking of the degeneracy with respect to $M$ \cite{becke,baerends}.
Once this degeneracy is broken, the corresponding ground state carries a 
current, just as in the present calculations. According to \cite{becke} the 
original degeneracy is restored if the current dependence of the $xc$ 
functional is explicitly accounted for.
The current-dependent functional used in Ref.~\cite{becke}
is not constructed within the CDFT of Vignale and Rasolt but 
based on analysis of the dependence of the curvature of the $xc$ hole on
the Kohn-Sham current and, as pointed out above, is not a vorticity 
functional. Implementation of the functional is done not self-consistently
(as in full CDFT) or perturbatively (as in pCDFT), but in a post-LDA manner,
in which orbitals obtained in a self-consistent (current-independent) LDA 
calculation are substituted once into the current-dependent functional
\cite{footnote2}.
In spite of these methodological differences between the present work and 
Ref.~\cite{becke}, Becke also finds that to
within the accuracy of today's density-functionals the current-dependent terms
cannot be neglected in high-precision DFT calculations of open-shell atoms.
In fact, as illustrated in the third column of Table \ref{table2}, the size 
of the current-dependent corrections to the total energy differences
obtained from subtracting the values in the last two columns of Table~2 of 
Ref.~\cite{becke} is rather similar to the size of the current-induced
corrections to $\epsilon_k$, calculated here.

\section{\label{conclusions}Conclusions}

We have performed CDFT calculations of ionization energies of open-shell
atoms prepared in a current-carrying state, with the aim to illustrate the
usefulness and viability of CDFT for electronic-structure calculations
in the presence of orbital currents. We summarize our conclusions as follows:

(i) The perturbative approximation to CDFT, pCDFT, has been implemented for 
the first time and was found to be numerically feasible. 

(ii) The CDFT $xc$ functional has been implemented in the linear-response
approximation (\ref{linres},\ref{ldakernel}) 
of Vignale and Rasolt \cite{vr1,vr2}. Three different 
parametrizations for the orbital susceptibility entering this functional
have been tested. The LCH 3-term parametrization was found to be slightly 
less accurate than claimed in the original reference 
(rms error $1.9\times 10^{-3}$ instead of $1.5\times 10^{-3}$). 
Two alternative parametrizations were developed. Our 3-term parametrization 
leads to an rms error of $1.2\times 10^{-3}$, while the 5-term 
expression we used in our numerical calculations reduces this to 
$2.1\times 10^{-4}$.

(iii) Orbital currents in open shells result in small but not negligible
shifts of the ionization energies. Such shifts can become important, e.g., when
ionization takes place in the presence of external electric or magnetic fields.
The same applies to the formation and breaking of chemical bonds in the
presence of such fields, and to the TDDFT calculation of excitation energies.
The calculated shifts, however, do not consistently improve agreement with 
experiments carried out in the absence of magnetic fields. Presumably this is 
due to the selection of specific current-carrying states in the calculation, 
which are averaged over experimentally. A more conclusive test of (p)CDFT 
would thus require experiments carried out in the presence of magnetic fields, 
giving rise to a well defined current in the open shell.

(iv) The order of magnitude of the pCDFT terms shows that modern 
density-functional techniques have reached an accuracy at 
which the magnitude of small current-related effects arising in open shells is
beginning to be significant compared to the remaining difference to experiment.
Hence further refinement of DFT-based calculations of atomic spectra should
consider the possibility of spontaneous or induced currents in open shells
\cite{handy3,krieger,jcpbaerends,becke}.

More research is needed to extend this analysis to other atoms and to 
molecules. Fully self-consistent (neither perturbative nor post-LDA) CDFT 
calculations would be desirable to this end, as is, obviously, construction 
of more reliable current-dependent $xc$ functionals.

{\bf Acknowledgments}\\
This work was sup\-por\-ted by FAPESP. One of us (KC) thanks E.~K.~U.~Gross, 
G.~Vignale and S.~Kurth for useful discussions, and A.~B.~F.~da~Silva for 
hospitality at the IQSC. We would like to thank E. Engel for providing his 
code OPMKS on which the numerical calculations of this work were based.

\end{document}